\documentclass[num-refs]{wiley-article}

\usepackage{siunitx}
\usepackage{setspace}
\usepackage{hyperref}

\papertype{Original Article}
\paperfield{JSM 2021}
\title{Assignment-Control Plots:\\ A Visual Companion for Causal Inference Study Design}

\abbrevs{IV, instrumental variable; CABG coronary artery bypass grafting}

\author[1, 2]{Rachael C. Aikens}
\author[3]{Michael Baiocchi}

\affil[1]{Program in Biomedical Informatics, Stanford University, California, USA}

\affil[2]{Department of Statistics, Stanford University, California, USA}

\affil[3]{Department of Epidemiology and Population Health, Stanford University, California, USA}

\corraddress{Michael Baiocchi, Redwood Building, 150 Governor's Lane, Stanford, CA 94305-5405, USA}
\corremail{baiocchi@stanford.edu}

\presentadd{Redwood Building, 150 Governor's Lane, Stanford, CA 94305-5405}

\fundinginfo{National Institutes of Health, Grant/Award Number: T32 LM012409; Stanford Graduate Fellowship in Science and Engineering}

\runningauthor{}

\begin{document}

\begin{frontmatter}
\maketitle

\begin{abstract}
An important step for any causal inference study design is understanding the distribution of the treated and control subjects in terms of measured baseline covariates.  However, not all baseline variation is equally important.  In the observational context, balancing on baseline variation summarized in a propensity score can help reduce bias due to self-selection. In both observational and experimental studies, controlling baseline variation associated with the expected outcomes can help increase the precision of causal effect estimates. We propose a set of visualizations which decompose the space of measured covariates into the different types of baseline variation important to the study design.  These ``assignment-control plots'' and variations thereof  visually illustrate core concepts of causal inference and suggest new directions for methodological research on study design. As a practical demonstration, we illustrate one application of assignment-control plots to a study of cardiothoracic surgery. While the family of visualization tools for studies of causality is relatively sparse, simple visual tools can be an asset to education, application, and methods development.

\end{abstract}
\end{frontmatter}

\section{Introduction} \label{intro}
A fundamental problem of causal inference is the impossibility of observing counterfactuals: Once an individual has received a treatment or exposure, we can never observe what might have happened to that individual had they not received treatment, and vice versa. This makes individual-level treatment effects difficult or impossible to observe directly. This problem is generally addressed by comparing the outcomes of treated individuals and untreated individuals in some setting which controls for ways that these two groups might systematically differ.  Intuitively, our observations of the control sample are used as approximations to help us understand what \textit{might} have happened to the treated individuals had they been untreated.  A question fundamental to these approaches is: How would we like the compared treated and control samples to be similar (or different) in order to obtain a clear understanding of the causal effect?

Researchers have proposed a variety of matching and conditioning methods to address baseline variation between treated and control samples in different observational contexts. Mahalanobis distance matching methods seek to match treated and control individuals which are similar with respect to \textit{all} measured covariates. However, not all measured covariates are necessarily important to the causal problem -- especially as researchers collect more and more comprehensive observational data with many measured covariates.  This begs the question: what aspects of baseline variation are most important to address in order to estimate the causal effect with minimal bias and variance? 

One popular approach is subclassification or adjustment on an estimated propensity score, which summarizes the measured baseline variation influencing the probability of assignment. Intuitively, propensity score methods  model the treatment assignment mechanism based on observed covariates, so that it can be adjusted for. Under suitable assumptions (including no \textit{unmeasured} confounding), matching exactly on the propensity score recapitulates a completely randomized controlled experiment, allowing for identification and unbiased estimation of treatment effect.  However, critics of propensity score matching note that the propensity score tends to neglect baseline variation which is less associated with treatment assignment but influential on the \textit{potential outcomes} of the study subjects, potentially resulting in unfavorably high variance and low statistical power. \cite{king2016propensity}. 

The less-commonly discussed prognostic score, formalized by Hansen \cite{hansen2008prognostic}, models the expected outcome of each subject under the control assignment, based on the observed covariates.  Interestingly, under suitable assumptions, balancing on the prognostic score results in a form of covariate balance which leads to unbiased estimation of the causal effect, analogous to the propensity score \cite{hansen2008prognostic}. A small but growing body of literature suggests that methods which match jointly on a prognostic score and a propensity score may be a favorable approach in some observational contexts, optimizing directly for propensity score balance and prognostic score balance \cite{leacy2014joint, antonelli2018doubly, aikens2020pilot}.

Love plots (standardized mean difference plots) and histograms allow the researcher to check the balances of individual covariates across treatment groups.  However, as researchers are empowered to collect data sets with ever greater numbers of measured covariates, it will become increasingly necessary to identify which aspects of baseline variation are most important to a causal question and which are not.  Assignment-control plots, introduced briefly by Aikens et al. \cite{aikens2020pilot}, reduce the dimensionality of the covariate space by visualizing each subject of an observational study in terms of their propensity and prognostic score. These are two (often interrelated) features which are directly relevant to observational studies of causality: propensity score similarity between compared individuals reduces bias \cite{rosenbaum1983central}, while prognostic score similarity between compared individuals reduces bias as well as variance and increases power in sensitivity analyses of unobserved confounding\cite{hansen2004full, leacy2014joint, antonelli2018doubly, aikens2020pilot}.  Observing these two aspects of baseline variation and how they relate to each other, we can illustrate core concepts of causal inference, which can facilitate scientific communication and education, or provide insights that lead to new methodological hypotheses.  Although application is a secondary focus of this work, these plots may also be useful in diagnosing underlying issues with a real data set, or assessing the quality of a matching or stratification process. In this paper, we summarize several uses of assignment-control plots as a conceptual and visualization tool, as well as possible extensions to instrumental variable study designs.  While most of the examples in this report are theoretical illustrations, we conclude with an applied example in which we apply assignment-control diagnostic plots to a study of cardiothoracic surgery. We suggest that assignment-control plots and variations thereof may be a useful practical and conceptual visualization tool in many branches of causal inference research. 
\section{Materials and Methods} \label{methods}

\subsection{Notation and Background} \label{methods:notation}
We adopt the Neyman-Rubin potential outcomes framework, in which a sample is described by $$\mathcal{D} = \{(X_i, T_i, Y_i)\}_{i = 1}^n,$$ where the triplet $(X_i, T_i, Y_i)$ describes an individual with measured covariates $X_i$, binary treatment assignment indicator $T_i$, and observed outcome $Y_i$.  We take $Y_i(T)$ to represent the potential outcome of individual $i$ under treatment assignment $T$. The fundamental problem of causal inference is that it is impossible to observe both potential outcomes, $Y_i(0)$ and $Y_i(1)$ for any individual.

The propensity score is defined as $e(X) = P(T=1|X)$.  The popularity of the propensity score in observational studies stems primarily from its use as a balancing score, i.e.
\begin{equation}
    T \bot X | e(X)
\end{equation}
That is, within level-sets of the propensity score, the treatment assignment is independent of the measured covariates. Under the assumption of strongly ignorable treatment assignment, exact matching on the propensity score allows for unbiased estimation of the treatment effect \cite{rosenbaum1983central}.

The prognostic score is defined by Hansen as any quantity $\Psi(X)$ such that 
\begin{equation}
    Y(0) \bot X | \Psi(X)
\end{equation}
In essence, a prognostic score is any function of the measured covariates which -- through conditioning -- induces independence between the potential outcome under the control assignment and the measured covariates.  It is thus, by definition, a balancing score as well.  Under regularity conditions analogous to those for the propensity score, conditioning on the prognostic score also allows for unbiased estimation of the treatment effect, as described in more detail by Hansen \cite{hansen2008prognostic}.  When $Y(0) | X$ follows a generalized linear model $\Psi(X) = E[Y(0)|X]$.  In the literature, the prognostic score is often treated more informally as the expected outcome under the control assignment given the observed covariates.

\subsection{Set-up} \label{methods:setup}

The demonstrations that follow depict several simulated data sets.  In keeping with \cite{aikens2020pilot}, the primary generative model for these is as follows:
\begin{align*}
X_i &\sim Normal(0, I_{10}) \\
T_i &\sim Bernoulli\left(\frac{1}{1 + exp(\phi(X_i))}\right) \\
Y_i(0) &= \Psi(X_i) + \epsilon_i \\
\epsilon_i &\sim N(0, \sigma^2),
\end{align*}
where $\phi(X)$ and $\Psi(X)$ represent the true propensity and prognostic score functions.  In general, these will be given by:
\begin{align*}
\phi(X_i) &= c_1 X_{i1} - c_0 \\
\Psi(X_i) &= \rho X_{i1} + \sqrt{(1 - \rho^2)} X_{i2},
\end{align*}
where $c_1$, $c_0$, $\sigma^2$ and $\rho$ are constants.  In particular, the form for the prognostic function above guarantees that $\rho = Corr(\phi(X), \Psi(X))$. Note that 10 baseline covariates are measured for each individual, but only $X_{i1}$ and $X_{i2}$ are important to either the outcome or the treatment assignment. We will also briefly suggest some other possible assignment-control plots  generated using different forms for $\phi$ and $\Psi$ (quadratic and discontinuous, figure 2).  The code for this project is available on github at \href{https://github.com/raikens1/RACplots}{https://github.com/raikens1/RACplots}.

\subsection{Fitting the Score Models}\label{methods:fitting}

In observational studies in practice, the propensity and prognostic score models are not known.  Conventionally, the propensity scores are often estimated from a logistic regression of the baseline covariates on treatment assignment, fit on the entire study sample.  Fitting the prognostic score may be somewhat more nuanced \cite{aikens2020pilot}.  First, since the prognostic model is meant to predict the outcome under the control assignment, the prognostic model is fit only on controls.  Thus, all prognostic score estimates on the treatment population are necessarily extrapolations.  Second, fitting the prognostic model on the entire control population raises concerns of overfitting \cite{hansen2004full,abadie2006large,antonelli2018doubly}.  To address these concerns and preserve the separation of the design and analysis phases of the study, Aikens et al.\cite{aikens2020pilot}, propose a ``pilot design'', in which a subset of the control individuals is selected and held aside for the purpose of fitting the prognostic model.  These controls -- comprising a ``pilot data set'' are then discarded, so that the observational units used to train the prognostic model are disjoint from the set used in the final analysis (the ``analysis set'').  The question of how to appropriately select the control observations for the pilot set is a difficult one, described in more detail elsewhere \cite{aikens2020pilot,aikens2020stratified}.

For conceptual clarity, the theoretical examples that follow bypass the problem of score estimation by using the ground-truth propensity and prognostic scores, as specified by our simulation set-up. In section \ref{results:applied}, we consider an applied example in cardiothoracic surgery in which the true propensity and prognostic score models are not known and must be estimated in order to create an assignment-control plot. For more discussion on the realities of fitting the score models, see \cite{aikens2020pilot}.

\subsection{Applied Example}

As a demonstration of assignment-control plots in practice, we consider an applied example comparing 30 day mortality between female coronary artery bypass grafting (CABG) patients with and without a female primary surgeon.  This work is included as a conceptual example; a more thorough consideration of this question might include more nuanced corrections and methods not considered here. Patient covariates and outcome information was extracted from medicare claims data for 1,155,903 CABG surgeries from 1998 to 2016. The gender of the primary surgeon was obtained from the National Plan and Provider Enumeration System records.  When more than one surgeon was involved in a procedure, the primary surgeon considered to be the one with the highest volume of prior surgeries.  One limitation of this study design was that patient information included only ``sex'' and provider information included only ``gender.''  In addition, many providers had missing gender information.

382,688 surgeries were performed on patients whose sex was recorded as female.  After excluding surgeries with missing outcome information (17 observations) or missing gender information for the primary surgeon (81,233 observations), a total of 301,438 surgeries remained.  To protect patient privacy, random noise was added to patient age and surgery year, and medicare qualification status, race, admission type, and admission day were shuffled within groups of patients with the same outcome and exposure.  We fit a logistic propensity score model from the entire data set of 301,438 surgeries.  The prognostic model was fit using a logistic lasso on a ``pilot'' set of 5\% of the controls (14,726 observations), leaving an analysis set of 286,712 surgeries for the remainder of the analysis. This size of a pilot set is probably unnecessarily large for most practical studies, but facilitates easy fitting of the prognostic score for demonstration. In particular, since the outcome is known to be quite rare (30-day mortality for CABG is less than 5\%)\cite{hansen201530}, fitting the prognostic score effectively for this particular outcome is a difficult task. A more formal consideration of the study question might consider more thorough curation and imputation of covariate information for the fitting of the prognostic and propensity score models, and more sophisticated modeling techniques for addressing the large covariate space.

Love plots were constructed using the RItools package (v0.1-17) \cite{RItoolsManual, RItoolspaper}, and the pilot set was extracted using the stratamatch package (v0.1.5) \cite{aikens2020pilot}.  The lasso prognostic model was fit using the glmnet package (v 4.0) \cite{friedman2010glmnet}. Matches were created using the optmatch package (0.9-13) \cite{hansen2006optmatch}.

\section{results}\label{results}

\subsection{Assignment-Control Plots and Design Diagnostics}\label{results:diagnostics}

It is relatively commonplace to consider the distribution of a propensity score among observational study subjects. Researchers are often cautioned to check that the treated and control groups overlap in propensity score and that no individuals have propensity scores of approximately 0 or 1, since these conditions are necessary (though not sufficient) to ensure that the treatment effect is identifiable and that some aspect of the treatment assignment is random. Less attention is often paid to the distribution of the prognostic score and the relationship between propensity and prognosis in the data set. However, the connection between the likelihood of treatment and the expected untreated outcome can contain important information not captured in examinations of the propensity score alone.

\begin{figure}[h]
\centering
\includegraphics[width=6in]{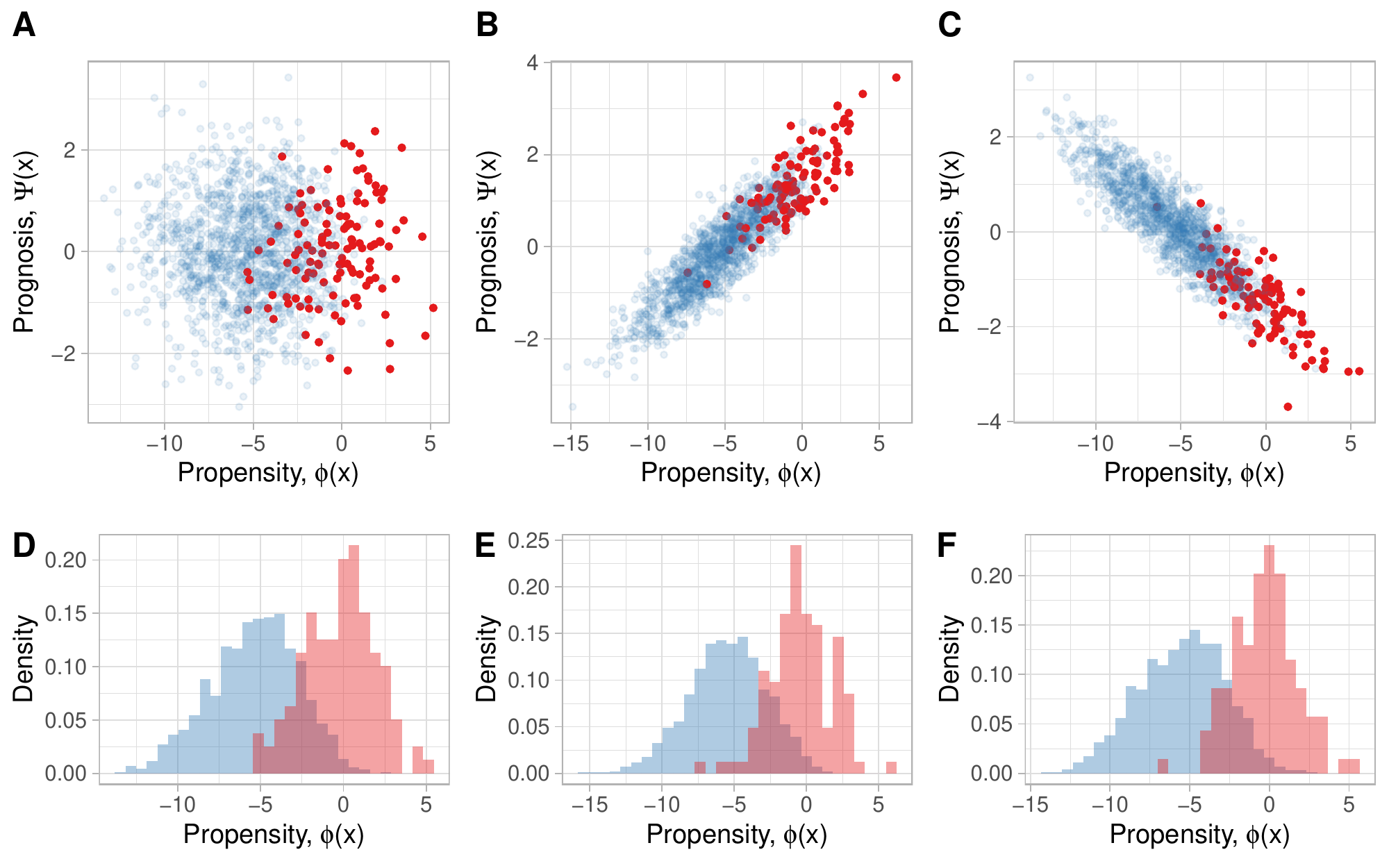}
\caption{Assignment-control plots (A-C) and propensity score density histograms (D-F) for three simulated observational data sets. Red points represent treated observations, blue points represent control.}
\end{figure}

Figure 1 shows example assignment-control plots (A-C) and propensity score density histograms (D-F) for three simulated observational data sets. Notably, the marginal distribution of the propensity score is identical across the three simulated scenarios, and the corresponding propensity score histograms (D-F) are qualitatively equivalent.  However, panels A-C reveal that the three settings are in fact quite different. In setting A, treated individuals are no different from control individuals in terms of their prognosis.  This means that even a naive comparison of treated and control subjects may give an unbiased treatment effect estimate under suitable conditions.  In settings B and C, treatment and control outcomes should not be directly compared, because the results would give a biased estimate of treatment effect.  Moreover, the direction of the correlation between propensity and prognosis suggests the direction of the bias from a naive comparison: If the individuals with the \textit{worst} prognosis are the \textit{least} likely to be treated (Figure 1B), we are more likely to \textit{overestimate} the effectiveness of the treatment in producing a more positive outcome.  If the individuals with the \textit{worst} prognosis are the \textit{most} likely to be treated (Figure 1C), we are liable to err in the \textit{opposite} direction.  Future work might consider whether this correlation indicates a tendency toward bias not only in naive comparisons, but for adjustment or subclassification approaches in which the score models are imperfect or matchings of treated and control individuals are not exact.

Another question evoked by figure 1 is the generalizability of the estimated treatment effect to different populations. For example, many matching studies focus on estimating the sample average treatment effect \textit{among the treated individuals}.  However, in scenarios B and C, the treatment groups and control groups are systematically quite different in terms of their prognostic score.  A researcher in this position should be prepared to ask: "Can a treatment effect estimated among the healthiest individuals in my sample generalize to the sickest individuals in a population?" and vise-versa.  Likewise, researchers seeking to estimate a sample average treatment effect or conditional average treatment effect might question whether such estimands can really be understood in settings in which treatment and prognosis are highly correlated, and overlap between the treated and control samples is sparse.  If few to none of the sampled treated individuals are among the sickest members of the population, this calls into question whether \textit{any} method -- however sophisticated -- can confidently estimate a treatment effect that applies to this group.

\begin{figure}[h]
\centering
\includegraphics[width=4.5in]{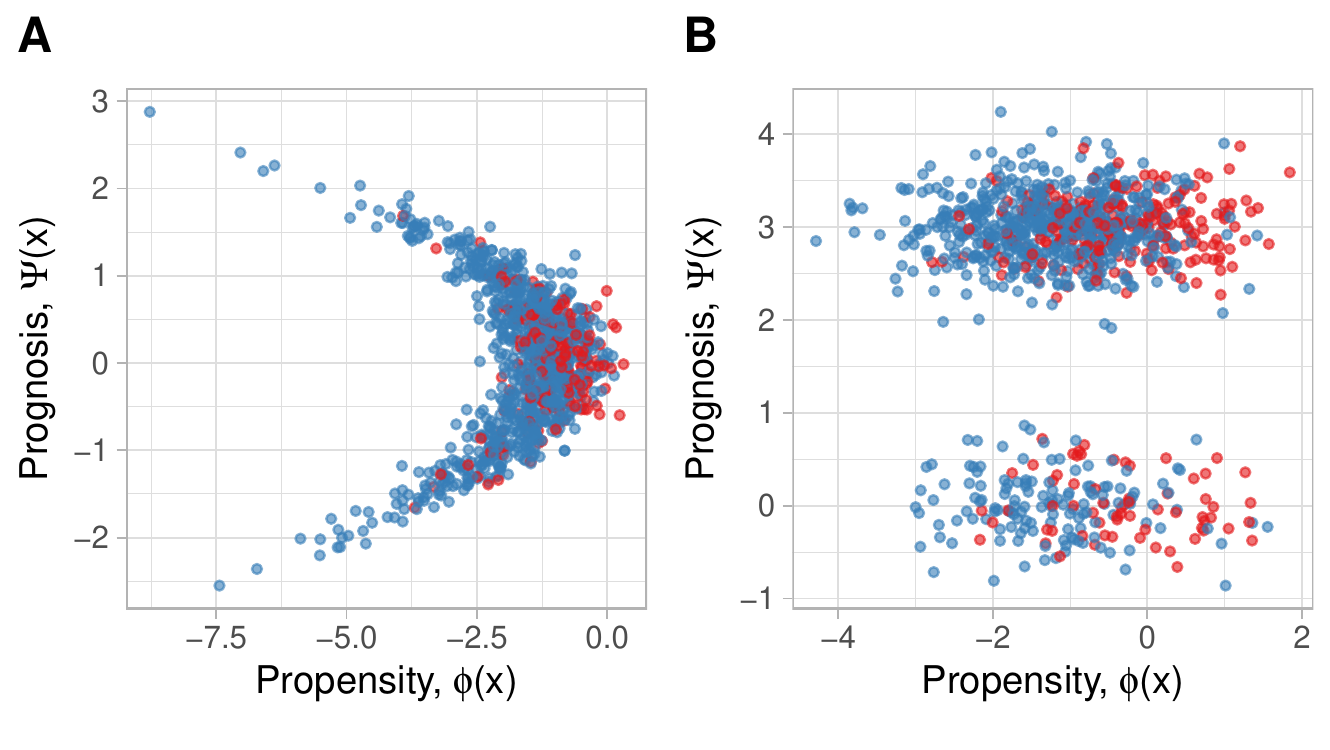}
\caption{Assignment-control plots of simulated data in two scenarios: (A) a discontinuity in the prognostic score (B) a nonlinear relationship between treatment and prognosis.}
\end{figure}

Figure 2 illustrates two other notable scenarios.  First, as suggested by figure 2A, a researcher need not assume that the relationship between propensity and prognosis is linear.  In the scenario shown, individuals with intermediate prognostic scores are the most likely to be treated.  This situation might arise with some frequency in medical settings: perhaps an intermediate treatment is rarely used for the sickest patients because they are thought to require more aggressive care; Or perhaps some patients have such poor prognosis that they are thought to be unlikely to benefit from treatment or to be at high risk for severe complications if treated. These scenarios suggest that there may be discernibly heterogeneous subgroups of individuals for whom treatment considerations differ, and careful thought is required to address whether the estimated treatment effect is likely to apply to different target populations. 

Second, while much attention is paid to the marginal distribution of the propensity score, an assignment-control plot can also be used to identify important patterns in the marginal distribution of the prognostic score.  Figure 2B represents a scenario in which some discrete covariate (e.g. sex, race, smoking status) has a profound correlation with the expected potential outcome under the control assignment, causing a strong separation in the prognostic scores. In scenarios like these, a researcher might consider stratifying or matching exactly within these groups in order to reduce prognostic variation between matched pairs. They might also consider whether this discrete covariate could be an important treatment effect modifier, or whether the mechanisms for prognosis and propensity are so different between these groups that they should be considered entirely separate samples with separate score models or even separate study designs. 

Interestingly, not every joint distribution of propensity and prognostic score is allowable on an assignment-control plot. In particular, if $\phi(X)$ is a true propensity score, then two points which have the same vertical position must have the same probability of treatment, by definition. In pictures, this means that the treated individuals \textit{must} be scattered evenly across vertical level-sets of the propensity score. The examples above suggest how several aspects of the design of a study and the interpretation of any results may be informed by thoughtful consideration of the possible joint distributions of propensity and prognostic scores. Other joint distributions may have different ramifications for study design and generalizability.

\begin{figure}[h]
\centering
\includegraphics[width=5in]{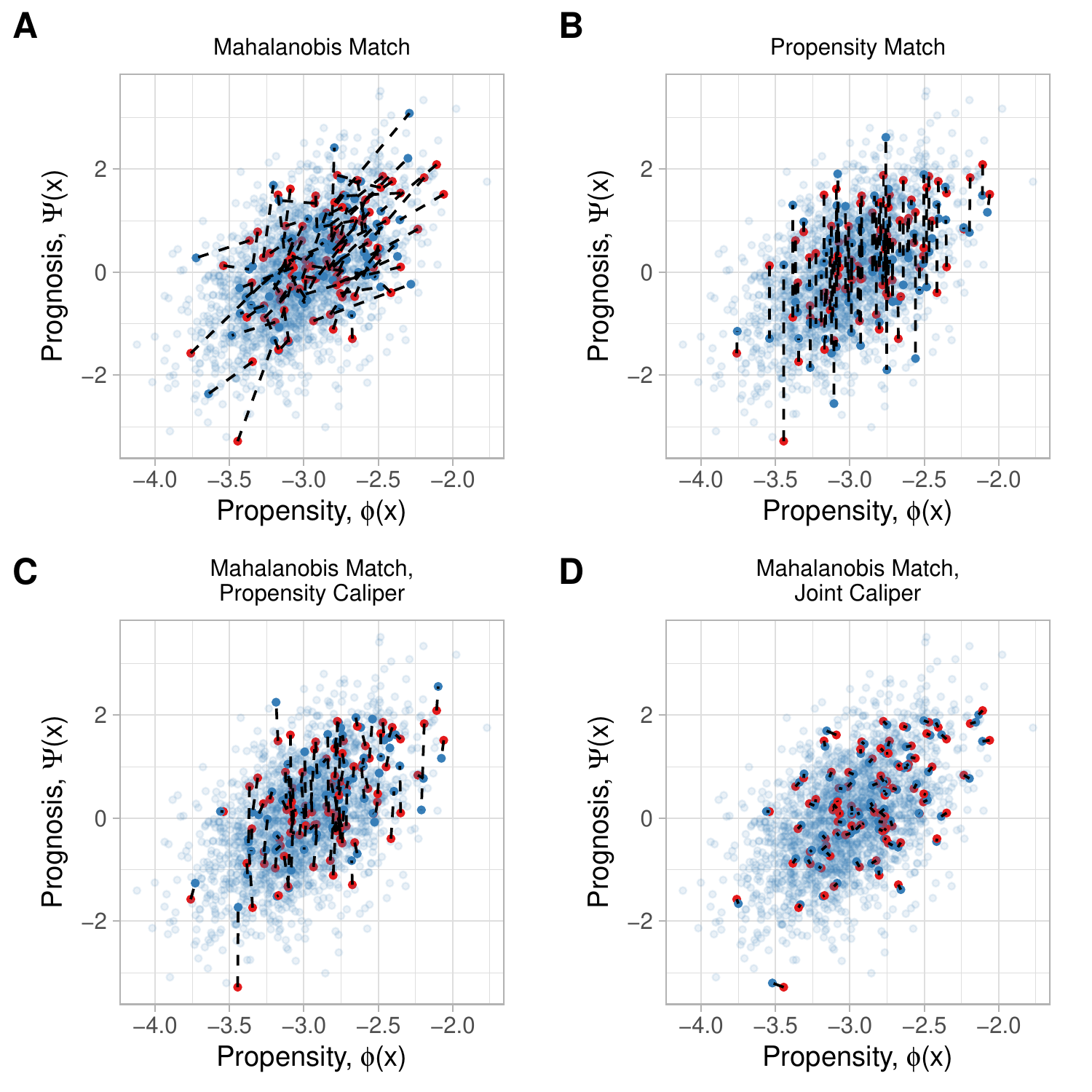}
\caption{Assignment-control plots depicting four different 1-to-1 matching schemes on the same simulated data set.  Red points represent treated observations, blue points represent control.  Dotted lines connect matched individuals. (A) Mahalanobis distance matching, (B), propensity score matching, (C) Mahalanobis distance matching with a propensity score caliper, (D) Mahalanobis distance matching with propensity and prognostic score calipers.}
\end{figure}

\subsection{Assignment-Control Plots and Matching}\label{results:matching}

Prior work suggests that jointly applying the propensity and prognostic score to matching studies in suitable scenarios may reduce variance and increase power in gamma sensitivity analyses, while increasing robustness in the case that one of the models is mis-specified (i.e. enabling doubly-robust estimation) \cite{aikens2020pilot, antonelli2018doubly, leacy2014joint}. These findings suggest that a desirable quality for an observational study design is that matched pairs are close together in assignment-control space. One appealing approach investigated by Leacy and Stuart \cite{leacy2014joint} is to match treated and control individuals based on Mahalanobis distance on the full covariate space, while enforcing calipers on the prognostic and propensity scores. Figure 3 visualizes this approach alongside common alternatives, illustrating several differences between these methods.

In Mahalanobis distance matching, all covariates are weighted equally in a statistical sense.  When there is an abundance of uninformative covariates (i.e. those which are not associated with the outcome, the treatment assignment, or any aspect of treatment effect heterogeneity.), Mahalanobis distance matching can select matches that may actually be quite distant in the assignment-control space (Figure 3A) \cite{aikens2020pilot}. On the other hand, propensity score matching optimizes directly for matches which are nearby in terms of the variation associated with the treatment (the "assignment" axis), but it is entirely agnostic to variation associated with the outcome (Figure 3B).  This can result in high variance in estimated treatment effect \cite{king2016propensity}.  Finally, the two caliper methods impose constraints on the matching process to ensure that matches are close in terms of propensity score (Figure 3C) or both propensity and prognostic score (Figure 3D). (Figure 3D) underscores why comparing prognostically similar treated and control individuals can make a study's results less easily explained away in gamma sensitivity analyses for unobserved confounding.  Intuitively, gamma sensitivity analyses imagines some ``unobserved confounding'' adversary who, with a strength of $\Gamma$, shifts the treatment probabilities of matched individuals in order to bias our results.  If our matched individuals are very close in terms of their likely outcomes, such an adversary can do less harm.

Assignment-control plots also visualize how matching quality can degrade when propensity score overlap between treated and control individuals is poor (Figure 4).  If there are some levels of the propensity score at which there are many treated individuals and few controls, the matching algorithm may have to reach very far away in assignment-control space in order to find adequate matches for the treated observations (Figure 4B).  When this happens, the matched control individual is likely to be not only distant in propensity score but systematically different in prognostic score, especially when propensity and prognosis are highly correlated. This systematic deviation can lead to bias in the effect estimate.

\begin{figure}[h]
\centering
\includegraphics[width=5in]{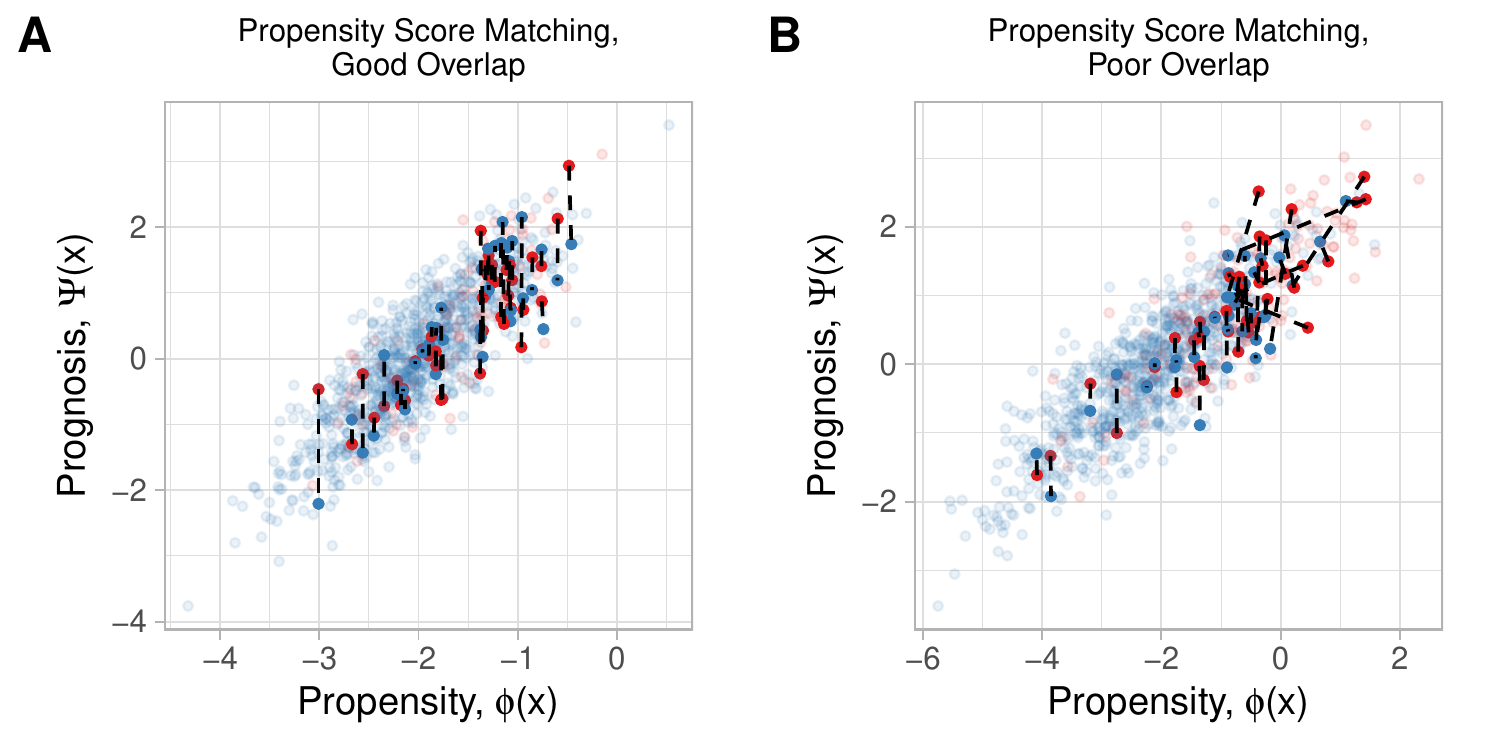}
\caption{Assignment-control plots of propensity score matches in scenarios with good (A) or poor (B) propensity score overlap.}
\end{figure}

\subsection{Assignment-Control Plots and Unmeasured Confounding} \label{results:SITA}

A wide and increasing variety of causal inference methods for observational studies -- in particular propensity score approaches -- depend on the absence of unobserved confounding. For matching studies, this dependence can be visually illustrated with assignment-control plots.

\begin{figure}[h]
\centering
\includegraphics[width=5in]{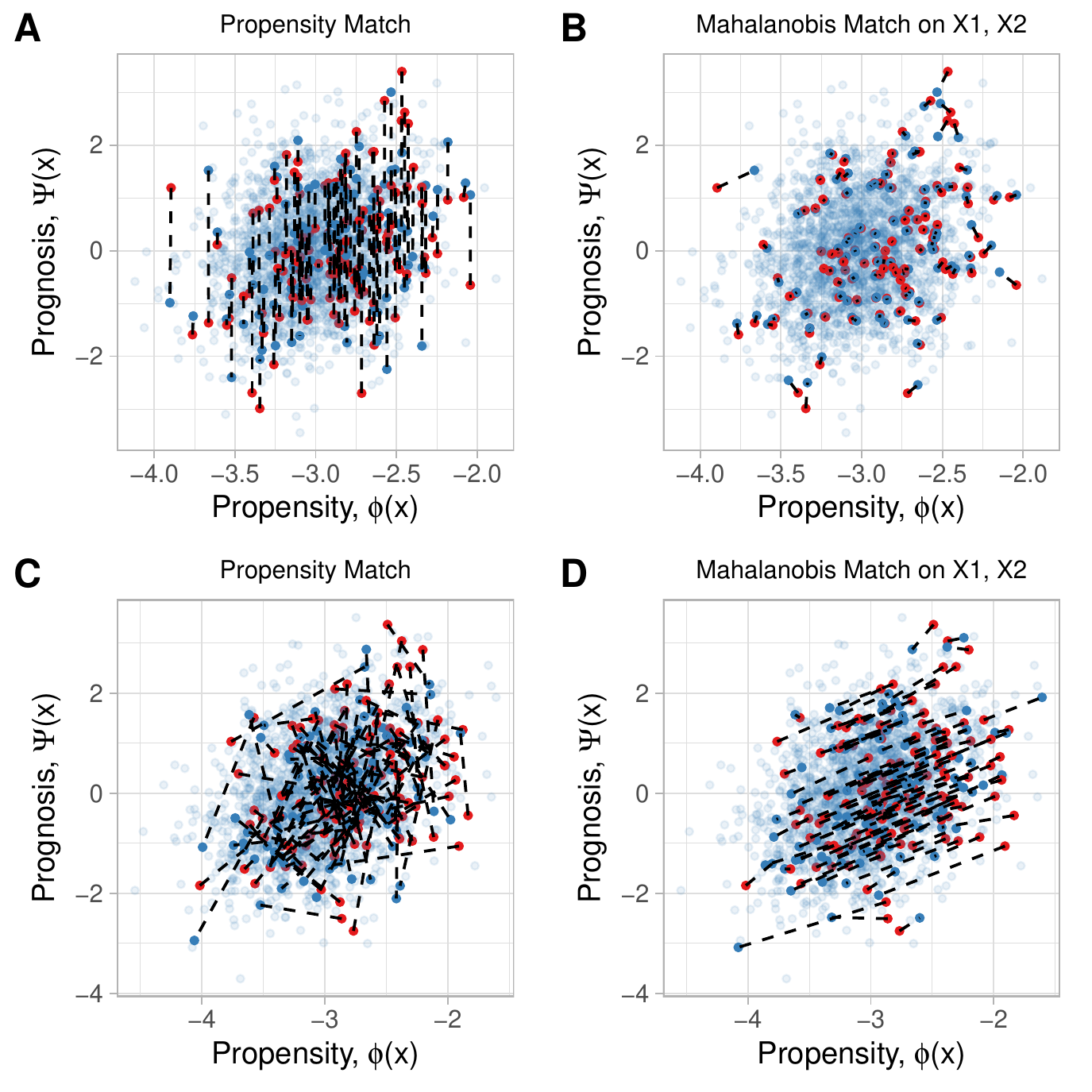}
\caption{Assignment-control plots for two matching schemes on a data set with unobserved confounding.  A-B depict the assignment-control space as ascertained without knowledge of the unobserved confounder.  C-D depict the true assignment-control space and the true match distances.}
\end{figure}

Figure 5 illustrates the behavior of two pair-matching approaches in a scenario with unobserved confounding.  We add to our data-generating set-up an unobserved confounder, $U$, such that:
\begin{align*}
    \phi(X_i) &= c_1 X_{i1} +  \eta U - c_0,\\
    \Psi(X_i) &=\rho X_{i1} + \sqrt{(1-\rho^2)}X_{i2} + \eta U,
\end{align*}

Where $\eta$ is a constant controlling the strength of $U$. Suppose we somehow ascertained exactly the correct relationships between the two score models and the \textit{observed} covariates, so that our propensity and prognostic models are precisely $\hat{\phi}(X_i) = c_1X_{i1} - c_0$ and $\hat{\Psi}(X_i) = \rho X_{i1} + \sqrt{(1-\rho^2)}X_{i2}$, respectively.  That is, the score models are exactly correct, except that they do not include the unobserved confounder. Figure 5 panels A-B depict the assignment-control plots we might make and matchings we might produce using these score models, $\hat{\phi}$ and $\hat{\Psi}$.  Since both the assignment-control plots and the matchings use only the observed covariates and exclude unmeasured confounding, propensity matches appear quite close in $\hat{\phi}$ (Figure 5B) and a Mahalanobis distance matching on the informative measured covariates ($X_1$ and $X_2$) appear quite close in the assignment-control space defined by $\hat{\phi} \times \hat{\Psi}$ (Figure 5C).  

However, panels C-D in Figure 5 show the same matches in the \textit{true} assignment-control space, in which $\phi$ and $\Psi$ are known to depend on the unobserved confounder, $U$.  In each matching, pairs tend to differ from each other due to baseline variations in the unobserved confounder which were not accounted for in the matching process.  The contrast between Figures 5B and 5D most cleanly illustrate how failing to account for $U$ results in systematic error: in the true assignment-control space, one matched individual in each pair tends to have both higher prognostic score and higher propensity score than its partner. Since this individual is more often the treated individual than the control individual, estimates of treatment effect based on this matching will tend to be biased. A similar narrative is true for the propensity score match, although the image is not as clear because there is a large amount of prognostic variation between matched pairs. Thus the unmeasured confounder, $U$, induces systematic differences between paired individuals, even after matching. This illustrates visually why a variety of matching approaches still produce biased results when unmeasured confounding is at play.

\subsection{Randomization-Control-Assignment Plots and Instrumental Variables} \label{results:RACplots}

Propensity and prognostic scores are by no means the only characterizations of baseline variation important to a causal question. Extensions of assignment-control plots might visualize some combination of propensity and prognostic score with other axes of variation important to a study design. Here, we consider one candidate: an additional axis summarizing instrumental variation. Briefly, an instrumental variable (IV) is a measured covariate which is associated with the treatment, but which has no effect on the outcome except through the treatment (for an introduction, see \cite{baiocchi2014instrumental}).  IV study designs rest on their own set of assumptions which require care and skepticism (namely, that a valid instrument can be isolated from measured baseline variation). However, unlike propensity score approaches, IV study designs \textit{do not} require the absence of unmeasured confounding, creating an interesting orthogonality between these methods.

Implicit in both instrumental variable and propensity score designs is the supposition that there are two components influencing each individual's treatment assignment: confounding variation and randomizing variation.  A subject's ``decision'' to be treated (or not) is directed by influences that are associated with their likely outcome (which we treat as ``confounders'') and influences which are unrelated to their outcome (which we treat as ``randomizers''). These components can be further broken down into measured and unmeasured variation. Informally,
\begin{equation*}
    \parbox{2cm}{\centering \linespread{0.8} \selectfont treatment assignment} \text{``=''} 
    \underbrace{\parbox{2.5cm}{\centering \linespread{0.8} \selectfont measured confounding variation}}_{\parbox{2.5cm}{\centering \linespread{0.8} \selectfont propensity score}} \text{``+''}
    \parbox{2.5cm}{\centering \linespread{0.8} \selectfont\textit{unmeasured confounding variation}} \text{``+''}
     \underbrace{\parbox{2.5cm}{\centering \linespread{0.8} \selectfont measured randomizing variation}}_{\parbox{2.5cm}{\centering \linespread{0.8} \selectfont IV}} \text{``+''}
    \parbox{2.5cm}{\centering \linespread{0.8} \selectfont\textit{unmeasured randomizing variation}}
\end{equation*} Randomizing variation is essential and desirable; it underlies the warrant for inference from randomized trials and well-designed observational studies. Confounding variation -- which is inherent to observational studies -- generally causes bias and must be addressed by any observational study design.   The propensity score seeks to control for measured confounding variation so that all that remains is the implicit random variation (unmeasured confounding variation is assumed to be absent).  IV designs seek to directly isolate the measured randomizing variation, so that confounding variation (measured and unmeasured) is balanced by the law of large numbers.  Much of the propensity score literature implicitly treats randomizing variation as unmeasured, whereas an explicit requirement of instrumental variable studies is that some randomizing variation can be isolated from the measured covariates.

Confounding and randomizing variation play distinct roles in an observational study, which motivates visualizing them separately and treating them differently. For instance, observations from theory and simulation suggest that IVs should \textit{not} be included in propensity score models -- even though they are associated with treatment assignment -- since this may actually \textit{increase} the bias and variance of the causal effect estimates in the absence of strong ignorability \cite{bhattacharya2007instrumental, wooldridge2009IV, myers2011IV}. As an illustrative example, consider the simulation set-up with unobserved confounding from section \ref{results:SITA}, except that now a new measured covariate $Z$, is present as an instrumental variable (IV):
\begin{align*}
    \phi(X_i, Z_i) &= c_1 X_{i1} + c_2 Z_{i} + \eta U - c_0,\\
    \Psi(X_i) &=\rho X_{i1} + \sqrt{(1-\rho^2)}X_{i2} + \eta U.
\end{align*}
where $c_2$, $c_1$, $c_0$, $\eta$ and $\rho$ are constants, and $U$ is an unmeasured confounder. Now, the ideal (somewhat modified) propensity score, $\tilde{\phi}$ would summarize just the variation in $X_1$ and omit any variation in $Z$.  In some ways this is a departure from the conventional description of the propensity score, in that $\tilde{\phi}$ should summarize confounding variation and \textit{exclude} randomizing variation (i.e. $Z$). Instead, $Z$ (the IV) should be summarized in its own "randomization" axis. Figure 6 visualizes of this "randomization-control-assignment" space, which -- for simplicity -- visualizes each individual projected down onto each pair of axes ($\tilde{\phi} \times \Psi$, $\Psi \times IV$, and $\tilde{\phi} \times IV$).  Like Figure 5C-D, these plots show the \textit{true} match distances in light of the unmeasured confounder, $U$.  

\begin{figure}[h]
\centering
\includegraphics[width=6in]{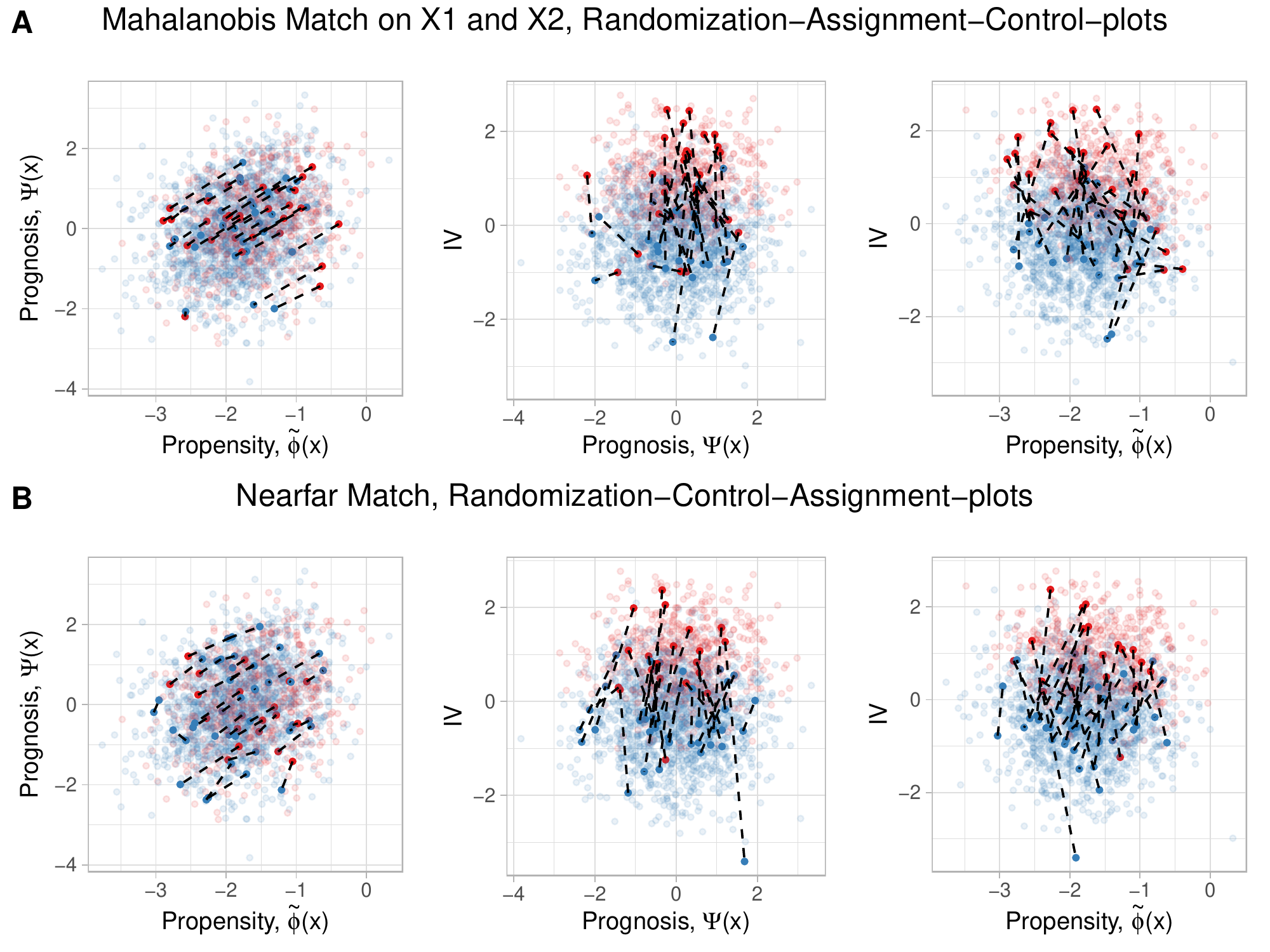}
\caption{Randomization-Control-Assignment plots.  Each panel in a trio depicts a different 2-D projection of the same data set within the randomization-control-assignment space.  Red points represent treated observations, blue points represent control.  Dotted lines connect matched pairs  (A) Depicts Mahalanobis distance matching on $X_1$ and $X_2$, while (B) depicts a nearfar matching of the same dataset. For visual clarity, only a subsample of 40 matches is shown.}
\end{figure}

Figure 6 considers randomization-assignment-control plots for two study designs.  The first (Figure 6A) performs Mahalanobis distance matching on just $X_1$ and $X_2$.  This matching pairs individuals based on measured confounding variation ($X_1$) and variation important to the potential outcome ($X_2$), ignoring the measured randomizing variation (the IV, $Z$).  The second design (Figure 6B) performs near-far matching, a matching design which directly uses the instrumental variable \cite{baiocchi2012near, baiocchi2014instrumental}.

The leftmost panel of Figure 6A is an assignment control plot.  This plot shows the unmeasured confounder, $U$, at work: although the measured confounding and prognostic variation has been efficiently matched upon, there is no way to ensure that matched individuals are close in terms of $U$.  This results in the ``slanting'' pattern previously shown in Figure 5D: since matched individuals differ in $U$ in a way that is beyond our control, one individual (the one with the higher $U$ value) will tend to be diagonal of the other.  This slant in the assignment-control space is for the most part unavoidable: when all of the measured confounding and prognostic variation is matched for, the contribution of unobserved confounders must remain.

How does randomizing variation, then, protect against the bias from unobserved confounding? Recall that there are two visual components to bias from unobserved confounding: (1) matched individuals are vertically distant from one another in the assignment-control plot (2) the treated individual is systematically more often the upper \footnote{or, for the opposite direction of bias, lower} individual of the pair.  Randomizing variation -- measured and unmeasured -- protects against unobserved confounding by disrupting component (2). In the rightmost panel of Figure 6A, the upper member of each matched pair is the treated individual more often than the control individual, a pattern which will cause bias in a treatment effect estimate from these matched pairs. This is due to the pressure imposed by the unobserved confounder, $U$. However, the upper member of the match is \textit{occasionally} the control individual rather than the treated one. This is possible because of randomizing variation, occasionally overcoming the pressure exerted by unmeasured confounding variation. In this way, the randomizing variation partially disrupts the tendency towards bias when an unobserved confounder is at work.  This is why including an IV in a propensity score -- or indeed, any matching scheme which seeks to \textit{minimize} IV distances within a matched set -- can be harmful: randomizing variation between individuals can actually be helpful, so removing it by matching or regression adjustment makes for worse estimation (supplementary figures 1-2).

The Mahalanobis distance matching example suggests how randomizing variation can be protective against bias even when it is ignored.  Well-used IV designs, however, can directly leverage measured randomizing variation to combat the influences of unobserved confounding.  Nearfar matching designs (figure 6B) are an IV approach which explicitly takes advantage of measured instrumental variation by pairing individuals who are \textit{near} in important non-instrumental covariates (figure 6B, left panel), but \textit{far} in terms of a measured instrument \cite{baiocchi2012near} (figure 6B, center and right panels). An important nuance to this study design is that paired individuals must be divergent in their IV but need not have the opposite treatment assignment.  Instead of comparing treated individuals to controls, the nearfar design compares those ``encouraged'' into the treatment group (i.e. by the IV) with those who were not \cite{baiocchi2010building, baiocchi2012near}.  This design directly disrupts the systematic tendency for treated individuals to deviate from matched controls by changing the criteria for who can be matched with whom in light of a measured IV.

With additional work, a visualization similar to Figure 6 may be an especially useful companion to nearfar matching designs \cite{baiocchi2012near}.  This may also be of use for other IV study designs for assessing the relationships between a candidate IV (or a set of weak IVs) and other covariates \footnote{Note: the process of fitting the propensity score may be nuanced in study designs using the IV, because selections of treatment assignment are "post-randomization" information in that they occur after the IV takes effect. Pilot design approaches may be a potential tool to avoid overfitting the propensity score in this scenario.}. The randomization-assignment-control plot may also have analogues in the randomized experiment setting, wherein the prognostic score is visualized alongside an axis (or axes) summarizing compliance behavior.

\subsection{Assignment-Control Plots in Application}\label{results:applied}

Much of the paper above is focused on assignment-control plots as a theoretical tool for communication and hypothesis generation. There is also ample potential for the assignment-control plot to be useful in applied studies - for example as data diagnostic (see section \ref{results:diagnostics}) or for assessing match quality (see section \ref{results:matching}). Here, we walk through an illustrative example, and discuss some considerations for using assignment-control plots in practice.

\subsubsection{Application to a Study of Cardiothoracic Surgery Outcomes}

Emerging evidence suggests that patient outcomes may improve when they are seen by clinicians whose identity (e.g. race, sex, gender) is similar to their own \cite{meghani2009patient, greenwood2018patient}.  In the field of cardiology, one ongoing concern is the observation that female patients tend to have worse outcomes for certain cardiovascular events and surgeries, particularly acute myocardial infarction. Recent work suggests that this disparity may be mediated in part by patient-physician gender concordance. Greenwood et al. \cite{greenwood2018patient} suggest that female patients may exhibit increased mortality from acute myocardial infarction when their physician is male. 

Observing that women also appear to exhibit higher operative mortality in coronary artery bypass grafting (CABG) surgeries \cite{blankstein2005female, mannacio2018sex}, we consider the question whether patient-physician concordance plays a role in 30-day mortality for CABG procedures.  In this illustrative example, we consider the comparison of 30 day mortality between female patients whose primary surgeons identified as female and those whose primary surgeons did not identify as female in a large data set of coronary artery bypass grafting (CABG) surgeries on Medicare patients from 1998 to 2016.  Note that a more thorough study of this question would need to address several alternative hypotheses, for example the possibility that differences in outcomes may be explained by different performances between male and female surgeons \cite{tsugawa2017comparison}.  Some of these nuances are discussed by Greenwood et al. \cite{greenwood2018patient}.

\begin{figure}[h]
\centering
\includegraphics[width=5.5in]{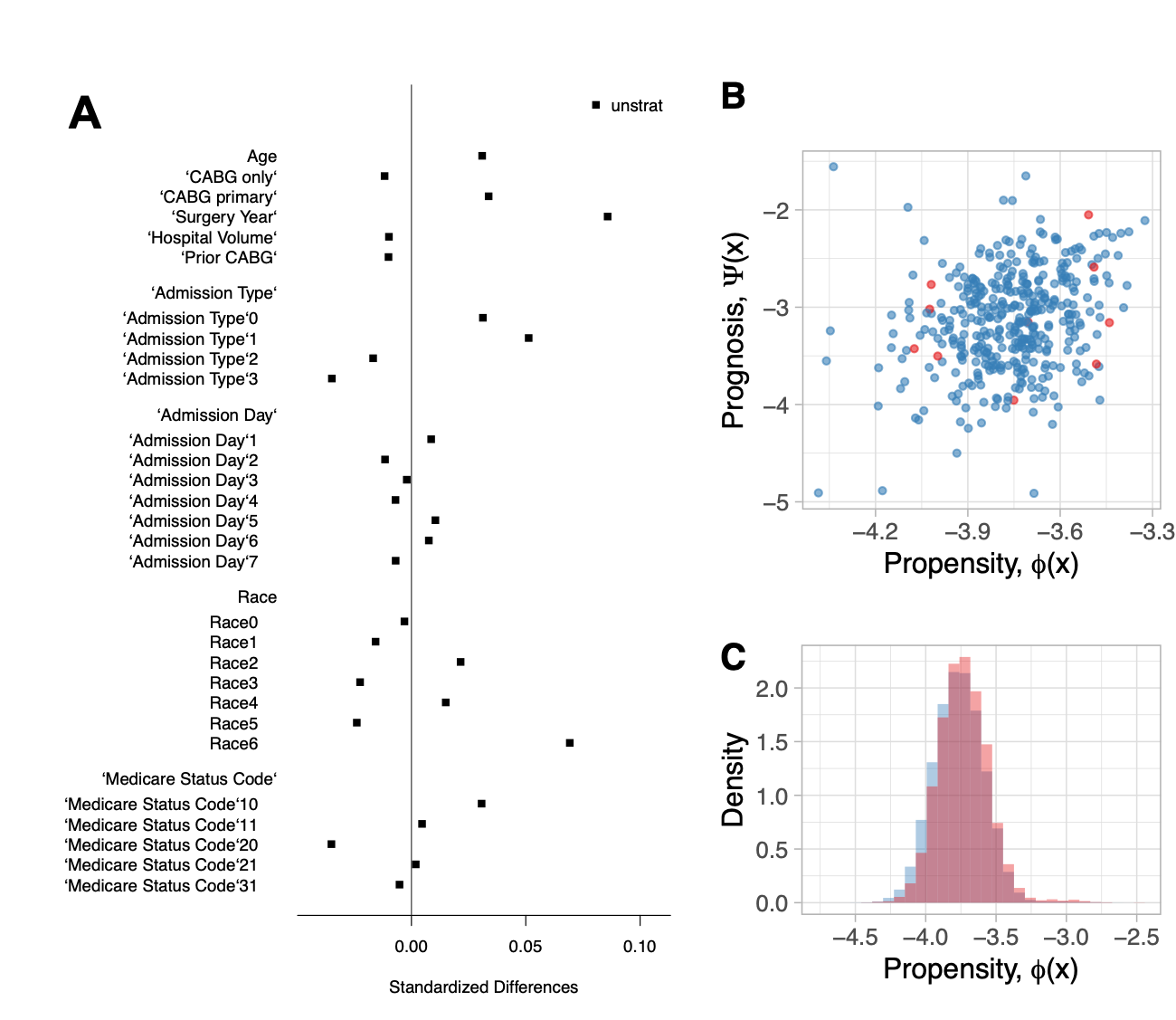}
\caption{Diagnostic plots comparing female CABG patients with and without female physicians as their primary surgeon.  (A) A Love plot, depicting mean differences in several baseline covariates between patients with female primary surgeons and patients without.  (B) An assignment-control plot for a random subsample of female patients. (C) A plot of propensity score overlap between the two populations. Red points and density represent patients with female surgeons, blue represents those without.}
\end{figure}

Figure 7A is a Love plot, depicting the standardized mean differences between ``treated'' (female primary surgeon) and control groups. This identifies some characteristics which tend to be more common in the female-surgeon group; for example these surgeries tend to occur in later years and on older patients.  This is an important consideration because 30 day mortality for CABG surgeries has improved substantially over the 18 years from 1998 to 2016 \cite{hansen201530}. We fit a logistic propensity score model on the full data set of 301,438 surgeries, predicting assignment to a female surgeon based on several covariates, such as the year of the surgery, day of the week, admission status code, and patient demographics.  Density histograms of the propensity score indicate that there is substantial overlap in propensity score between patients with female surgeons and those without (Figure 7C). 

Figure 7B shows an assignment-control plot for a random subset of the data set, with prognostic scores estimated using a pilot design.  Since only 6944 surgeries (2.36\%) of all surgeries on female patients were performed with a woman as the primary surgeon, ``control'' units were much more abundant than treated units.  This presents an ideal scenario for the use of a pilot design, since there are ample control units to select for the pilot set. We randomly subsampled 5\% of the control observations (14,726 surgeries) for the pilot set, stratifying by the home state of the patient. A prognostic model was fit the pilot set using a logistic lasso, and this was used to estimate the prognostic scores for the remaining observations.  A quick examination of the assignment-control plot suggests that prognosis and propensity are not highly correlated; that is, the variation most predictive of the outcome tends not to be the variation most predictive of the treatment assignment.  This is encouraging, because it is in agreement with the supposition that female surgeons are not systematically assigned to the most unhealthy patients nor the most healthy ones.

Figure 8 shows an example of two diagnostic plots for a matched subsample of 2,000 surgeries from the data set.  The Love plot (Figure 8A) compares standardized mean differences between the treated groups before and after matching.  Using this plot, it is easy to identify covariates which are poorly balanced in this matching, however, the researcher must use their judgement to decide which imbalances are important.  For example, it may be productive to match more closely on surgery year since we know this variable is associated with assignment to a female surgeon and with 30-day mortality, but imbalances in the day of the week of hospital admission may be less concerning.  The assignment-control plot gives a quick visualization of where each match falls in terms of propensity and prognostic score difference.  A researcher might consider applying calipers to obtain closer matches in the assignment-control space.

\begin{figure}[h]
\centering
\includegraphics[width=6in]{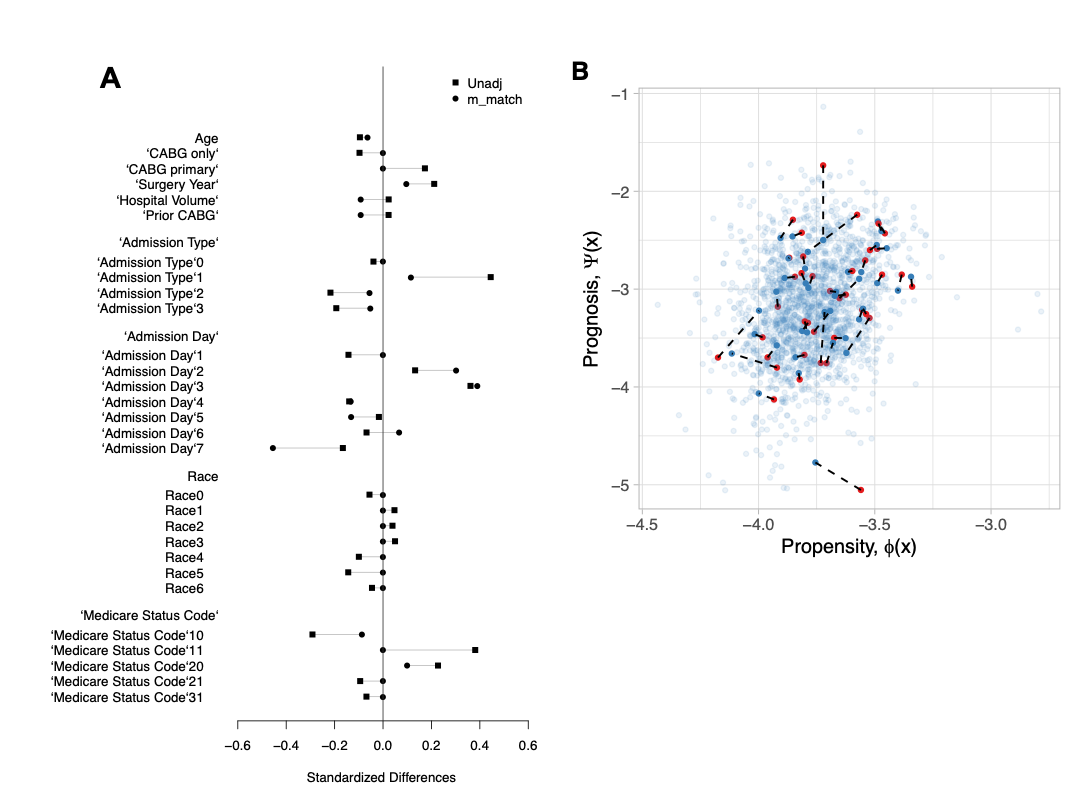}
\caption{Diagnostic plots for a Mahalanobis distance matching of a subsample of 2,000 CABG surgeries with and without female primary surgeons. (A) A Love plot comparing the unadjusted sample (black square) with the matched subsample (black circle). (B) An assignment-control plot depicting the matched pairs.}
\end{figure}

\subsubsection{Considerations for Assignment-Control Plots in Application}

A combination of Love plots and assignment-control plots may be a valuable diagnostic pairing for observational studies which use matching or subclassification.  Love plots are informative in that they allow the researcher to consider each covariate separately and quickly identify imbalances in especially important individual covariates.  However, Love plots do not directly convey any information about which covariates are most important to prognosis, which may make it difficult to quickly assess which covariate imbalances are worth extra attention when many covariates are at play. Assignment-control plots offer a quick way to gain intuition about the relationship between propensity and prognosis within a data set, allowing the researcher to identify potential violations of assumptions and to assess match quality in terms of propensity and prognosis simultaneously.

However, assignment-control in application plots suffer from many of the same weaknesses as propensity and prognostic score methods in general.  First, an assignment-control plot is only as reliable as the propensity and prognostic scores from which it is constructed.  If model fit to the analysis set is poor, or if there is unobserved confounding at play (section  \ref{results:SITA}), assignment-control plots may be misleading. Future work might consider how uncertainty bounds around propensity and prognostic scores might be estimated, displayed, and interpreted. Second, when using the prognostic score, it is advisable to hold aside a subsample of the control data for estimating the prognostic model in order to avoid overfitting. Aikens et al \cite{aikens2020pilot} discuss some considerations for when a ``Pilot Design'' such as this is most useful, and when it may be too costly of a data sacrifice.  

Our general advice for applying assignment-control plots this: If the decision has been made to use a prognostic score in the study design, there is low cost and potential benefit to generating assignment-control plots.  First, no additional data sacrifice is necessary to make assignment-control plots if the prognostic score was already in the study design to begin with.  Second, a researcher who has selected a study design which uses prognostic and propensity scores is already making the implicit assumption that the estimated scores are adequate, and checking an assignment-control plot can be a responsible way ensure that nothing anomalous or concerning appears in the marginal or joint distribution of the scores.  Assignment-control plots may also be useful in scenarios where propensity and prognostic scores are not a planned part of study design, although this requires more judgement regarding whether the data sacrifice to build a prognostic score solely for visualization is a worthwhile trade-off for potential insights gained from the plot.

\section{Discussion}

Assignment-control plots are nested in a broader conversation about the differing types of baseline variation and their differing significances to a causal question. The ways these sources of variation interact and how they can be leveraged or mitigated against comprise  an implicit focal point of causal inference research, both methodological and applied.  The greater clarity our community has in defining and characterizing these concepts, the more effectively we can communicate, teach, design studies, and generate new insights or hypotheses.

A modern shift towards an emphasis on large, passively collected data sets presents a host of challenges and opportunities for researchers interested in causality.  As we collect wider data sets with more measured covariates, it will be increasingly important to prioritize the baseline variation that is most important to the causal question - correctly leveraging measured covariates which are useful, and ignoring measured covariates which are uninformative. The complementary tools of the propensity and prognostic scores are well-suited to aid in this endeavor because they summarize two important aspects of baseline variation in the measured covariates: variation associated with the assignment mechanism, and variation associated with the potential outcomes. Indeed, the central insight of doubly robust estimation is the joint use of the propensity and prognostic score. 

However, propensity and prognostic scores are not the only important sources of variation in a causal inference study.  Other study designs may depend on an instrumental variable, a forcing variable for regression discontinuity, a summary of baseline variation associated with treatment effect heterogeneity, or compliance information in an encouragement design.  We consider one extension of the assignment-control plot which adds a `randomization' axis, and suggest how this conceptual tool might be useful in studies which leverage an instrumental variable.  Assignment-control plots and variations thereof can be thought of as dimensionality reduction tools in that they digest a possibly very large covariate space into a meaningful reduced space that is easier to use and understand.  While the possible variations on this theme are numerous, they are fundamentally driven by the same insight: a principled understanding of the different types of baseline variation and their differing significances to a causal question can enable researchers to improve the design of causal inference studies and clarify the way we communicate about them.

\section*{Acknowledgements}
The authors would like to thank Dr. Jonathan H. Chen for his support and mentorship, and Dr. Guillaume Basse for helpful thoughts on this work.  R.C.A. is supported by the National Institutes of Health under grant T32 LM012409, and a Stanford Graduate Fellowship in Science and Engineering.

\section*{Conflict of Interest}
The authors have no conflicts of interest to disclose.

\printendnotes

\bibliography{citations}

\end{document}